# High yield production of ultrathin fibroid semiconducting nanowire of $Ta_2Pd_3Se_8$

Xue Liu,[†,‡] Sheng Liu,[‡] Liubov Yu. Antipina,[§,∥,⊥] Yibo Zhu,[Δ] Jinliang Ning,[†] Jinyu Liu,[†] Chunlei Yue,[†] Abin Joshy,[†] Yu Zhu,[+] Jianwei Sun,[†] Ana M Sanchez,[¶] Pavel B. Sorokin[§,∥], Zhiqiang Mao,[†] Qihua Xiong,[‡] and Jiang Wei*[,†]

[†] *Department of Physics and Engineering Physics, Tulane University, New Orleans, Louisiana 70118, United States*

[‡] *Division of Physics and Applied Physics, Nanyang Technological University, 637371, Singapore*

[§] *National University of Science and Technology "MISiS", Moscow 119049, Russian Federation*

[+] *Department of Polymer Science, The University of Akron, Akron, Ohio 44325, United States*

[∥] *Technological Institute for Superhard and Novel Carbon Materials, Moscow 108840, Russian Federation*

[⊥] *Moscow Institute of Physics and Technology, Dolgoprudny 141700, Russian Federation*

[Δ] *Department of Mechanical Engineering, Columbia University, New York, New York 10027, United States*

[¶] *Department of Physics, University of Warwick, Coventry, CV4 7AL, United Kingdom*

*Corresponding to <u>jwei1@tulane.edu</u>

ABSTRACT

Immediately after the demonstration of the high-quality electronic properties in various two dimensional (2D) van der Waals (vdW) crystals fabricated with mechanical exfoliation, many methods have been reported to explore and control large scale fabrications. Comparing with recent advancement in fabricating 2D atomic layered crystals, large scale production of one dimensional (1D) nanowires with thickness approaching molecular or atomic level still remains stagnant. Here, we demonstrate the high yield production of a 1D vdW material, semiconducting $Ta_2Pd_3Se_8$ nanowires, by means of liquid-phase exfoliation. The thinnest nanowire we have readily achieved is around 1 nm, corresponding to a bundle of one or two molecular ribbons. Transmission electron microscopy and transport measurements reveal the as-fabricated $Ta_2Pd_3Se_8$ nanowires exhibit unexpected high crystallinity and chemical stability. Our low frequency Raman spectroscopy reveals clear evidence of the existing of weak inter-ribbon bindings. The fabricated nanowire transistors exhibit high switching performance and promising applications for photodetectors.

Van der Waals (vdW) two dimensional (2D) materials, such as graphene,[1-3] hexagonal boron nitride (h-BN),[4] transition metal dichalcogenides (TMDCs)[5-9] and black phosphorus[10] have attracted enormous research attention due to their extraordinary properties after thinning down to atomic level. Such high quality mono or a-few-layer can be achieved with simple mechanical exfoliation, in which adhesive tape is used to effectively separate layers that are originally bonded by weak vdW interactions.[1,3] The high crystalline films enable comprehensive studies of their novel quantum phenomena. Hidden behind the success of achieving these well preserved 2D film materials is the critical role of Van der Waals bond, weakly holding the layers together, but also chemically insert enough acting as a protective barrier to prevent degradation caused by the surface reaction. Moreover, the idea of mechanical exfoliation of 2D vdW material can be extended to one dimensional (1D) fibroid crystals, which consist of unit 1D molecular ribbons parallel bonded by vdW forces.[11-17] This presents a new path of fabricating TMDC nanoribbons and avoids the edge problems faced by traditional methods.[18-21] in which nanoribbons are directly cut out from the 2D TMDC layers. Indeed, mechanical exfoliation has been successfully applied to exfoliate the vdW fibroid bulk crystal - $Ta_2Pd_3Se_8$ (TPS), which can be viewed as an assemble of parallel running zigzag molecular ribbon (also we called it chain in the text below)[11] with edges terminated by Pd atoms and bonded together with interchain vdW bonding. Nanowires composed of a few TPS molecular chains as thick as a few nanometers has been achieved, demonstrating the feasibility of easy access to ultrathin high quality 1D nanowires for studying their intrinsic properties as well as potential applications.

However, mechanical exfoliation method is limited only in small laboratory experiments due to the extremely low productivity. To produce large quantities or scales of nanostructures for more practical applications, researchers have made efforts to discover alternative methods. One typical technique, known as liquid-phase exfoliation (LPE), has been proposed to break vdW bonding by utilizing ultra-sonication assisted by chemical solvents. It has been demonstrated to be a scalable and effective method producing a large number of vdW nano-flakes from their bulk counterpart as mentioned,[22-29] and carbon nanotubes from its bundles.[30-33] More interestingly, vdW interactions are widely existing not only in 2D vdW crystals but also in fibroid 1D/quasi-1D vdW crystal systems. Broad fabrication of 1D nanowires is expected to be realized by extending these exfoliation methods from 2D to 1D vdW crystal systems.

Here we report on our LPE study of single crystal 1D semiconducting TPS fibroid nanowires. In particular, we find that by using the LPE method ultrathin TPS nanowires as thin as 1 nm to 4 nm can be efficiently produced. Our comprehensive studies through low frequency Raman spectroscopy and high-resolution scanning transmission electron microscopy (STEM) demonstrate highly preserved crystal quality and weak inter-ribbon bindings. Then, the 1D nanowire transistors fabricated from liquid exfoliated TPS show promising potential in electrical switching applications, and a photothermal dominated photoresponse mechanism has also been revealed. Our work demonstrates a successful large-quantity fabrication of ultra-thin fibroid nanowires through a simple and clean technique. At this early stage, although our semiconducting TPS nanowires exhibit relative low mobility when compared to up-to-date record best carbon nanotubes[34-37] and silicon nanowires,[38,39] our fabrication method avoids the complexity caused by selection, separation, and protection problems remaining in current 1D system studies.[40] Besides, our STEM and transport results simultaneously indicate the outstanding stability of TPS nanostructures even down to the nanometer scale, which is crucial for next generation modern electronics.

## RESULTS AND DISCUSSION

**Exfoliations and basic characterizations**

$Ta_2Pd_3Se_8$ belongs to the isostructure group of ternary chalcogenides $M_2X_3Y_8$ (M = Ta, Nb; X = Ni, Pd, Pt; Y = S, Se), which has a linear crystal structure with bonding strengths differ near 20 times between and within unit "ribbons" (the smallest repeating unit in crystal structure as highlighted by the dashed rectangular in Figure 1a.[11] For each ribbon, it contains two chains of edge-sharing Se trigonal prims centered by Ta atoms and bridged by Pd atoms. The chains are further capped by additional Pd atoms at the two sides. The lateral dimensions of one unit ribbon is 1.0 nm × 0.4 nm. The ribbons are interconnected via weak bindings between terminated-Pd (trans-Se) atoms and trans-Se (terminated-Pd) atoms with neighboring ribbons, forming a windmill shape framework extending along *c*-axis. From the crystal structure, we believe that the existence of isotropic vdW bonds in TPS crystal is crucial for successful isolation of ultrathin nanostructures or even single molecular chain out from the bulk counterparts. To perform the LPE, the sonication power is applied to effectively break the inter-ribbon interactions so as to obtain separated thin nanowires, as illustrated by Figure 1b. To process the exfoliation, we surveyed different solvents including isopropyl alcohol (IPA), N-methyl-2-pyrrolidone (NMP), chloroform, etc (see Supplementary Table 1). The bulk TPS crystal of 0.5 mg was immersed into different solvents (20 ml) and then ultra-sonicated for 4 hours at 50ºC with a total input power of 0.6 MJ. After analyzing the exfoliation results, we note that NMP and IPA were the most appropriate solvents resulting in TPS nanowires dispersions as shown in Figure 1c, and NMP mixture was more stable without significant reaggregation after a few hours (see Supplementary Figure 1). The sonicated mixtures were then centrifuged at 500 rpm for 1 hour, the sediment was weight as ~21 wt% of the initial bulk material, and the supernatants were collected as the resultant liquids indicated by Figure 1d. We performed further characterizations based on deposited nanowires on substrate ($SiO_2$/Si wafer), referred to as on-chip sample (see Methods). Figure 1e and 1f show the as-prepared samples (NMP solvent) under Scanning Electron Microscopy (SEM) before and after centrifugation respectively. Before centrifugation, the TPS nanowires were densely overlapped on the wafer; while after centrifugation, individual nanowires with random orientations can be easily found.

From our observation, there is no significant un-separated or aggregated big chuck of crystals after exfoliations, we believe that, the bulk TPS has been extensively separated to give nanowires with certain thickness and length distributions. Then, we systematically investigated the thickness and length distributions of the nanowires produced by IPA and NMP solvents using Atomic Force Microscopy (AFM). As shown in Figure 2a and 2b, a variable range of thicknesses and lengths of TPS nanowires are demonstrated by the AFM images of randomly selected areas from IPA and NMP samples, respectively. Zoom-in views of 5×2.5 μm$^2$ squares indicate TPS nanowires as thin as 1.4 nm can be readily achieved. It is worth noting that the thinnest nanowire observed from our AFM is ~1 nm (Supplementary Figure 3), which should correspond to nanostructures consisting of 1-2 unit molecular chains as compared with the simulation (see Supplementary Figure 2). To better compare the results, the contour plots of the histograms of the nanowire thickness and length distributions, as obtained from 400 individual nanowires, are shown in Figure 2c and 2d for IPA and NMP exfoliations respectively. Both of the IPA and NMP exfoliated nanowires have thicknesses that are most distributed (> 90% number fraction) below 10 nm, while the NMP exfoliation yields slightly thinner and shorter nanowires in average, for which we can achieve ~45.57% nanowires below 4 nm. We also estimated the mass fractions of different thicknesses after centrifugation (see Methods). The mass fraction for nanowires below 10 nm is 48.3 wt%, and 4.8 wt% for nanowires below 4 nm. Thus the total yield from the initial material can be estimated as 38.2 % for nanowires below 10 nm and 3.8% below 4 nm. Our following discussions are all based on NMP prepared sample.

**Low-wavenumber Raman spectroscopy and transmission electron microscopy studies**

Based on the on-chip deposited nanowires, we performed low-wavenumber Raman spectroscopy study (see Methods). Vibrational modes located at low-wavenumber range less than 100 cm$^{-1}$ can denote the movement of atoms or molecules governed by weak forces, including vdW, quasi-vdW and hydrogen bond. Low-wavenumber Raman spectroscopy has been successfully used to investigate the interlayer phonon modes in 2D materials[41] and the intermolecular interactions between aromatic molecules.[42-44] Our detailed

Raman spectroscopy studies provide a guidance for further understanding of one-dimensional crystal structure and symmetries. As shown in Figure 2e, Raman spectra at different polarization configurations for a relatively thick nanowire (diameter ~1 μm), have been stacking plotted together. We measured the excitation-angle dependences of the Raman peaks, where 0º and 90º represent parallel ($\bar{z}(xx)z$) and cross ($\bar{z}(yx)z$) polarization configurations. A few peaks in the low wavenumber region (<100 cm$^{-1}$) are examined under different configurations, i.e. 18.1 cm$^{-1}$, 26.9 cm$^{-1}$, 32.5 cm$^{-1}$, 60 cm$^{-1}$, and 72.9 cm$^{-1}$ can be clearly resolved at $\bar{z}(yx)z$ configuration. The successful demonstrations of low wavenumber Raman peaks indicate the existence of inter-chain vibrations in our TPS system. In addition, these measured Raman peaks exhibit distinct excitation-angle dependent polarizations, as demonstrated by the intensity polarization angle dependences in Figure 2f. Here, we take P1 (32.5cm$^{-1}$), P2 (72.9cm$^{-1}$), and P3 (201.5cm$^{-1}$) as the examples of these different polarizations. More specifically, the Raman intensity of P1 is consistent with the laser polarization which reaches maximum at parallel configuration and minimum at cross configuration, while P2 intensity shows the opposite evolution and P3 intensity stays almost constant. These different polarization dependences indicate distinguishable symmetries related to these Raman modes.[41,45] Since the detector polarization is aligned with our TPS chain direction, vibrations along the chain and totally symmetric vibration modes will preserve the laser polarization in Raman scattering. Therefore, we suggest that the P1 mode is breathing-like vibration between chains or shearing-like vibration along the chain, and P2 mode is shearing-like vibration perpendicular to the chain. P3 mode shows a slightly broader peak which may correspond to a combination of two vibrations with close energy, one is mainly along the chain direction and another shows significant component perpendicular to the chain. We have simulated the possible corresponded atoms vibration modes, as shown in Supplementary Figure 4 and Figure 5. However, due to the complexity of the crystal structures, we are not able to correlate the irreducible representations for these peaks, which is beyond the scope of this research work. In fact, we also performed simulation of Raman spectra evolution with different thicknesses, as can be seen in Supplementary Figure 6, only single and double ribbons produce unique spectra. For example, in Raman spectrum of single ribbon, the modes corresponded to vibration of the bonds between ribbons are absent. With increasing nanowire size, its Raman peaks tend to the bulk spectrum.

To further analyze the crystallinity of the exfoliated TPS nanowires, we preformed aberration-corrected scanning transmission electron microscopy (STEM) characterization. High quality nanowires with various thicknesses can be easily identified. Figure 3a shows annular dark-field (ADF) images of a thicker (9.3 nm) TPS nanowire along the <100> zone axis. The orientation can be corroborated by the fast Fourier transform (FFT) as indicated by the top right inset. The interplanar distances correspond with the (100), (2-11), and (2-10) planes (compared with the result of crystal structure simulation). The atomic resolution image of the TPS nanowire clearly shows a smooth and clean surface without significant degradations, as shown by bottom right inset of Figure 3a. Figure 3b shows the ADF images of an ultra-thin (~3.0 nm) TPS nanowire. The nanowire shows a few segments orientated in different directions as highlighted in the image. At a higher magnification, we observed atomic resolution images of this specific nanowire at three selected areas. The crystal orientation can be identified by comparing to the simulated crystal structure with six unit chains as shown by the Figure 3b inset. The view directions of area 1 and area 2 are [120] and [100] respectively. Area 3 shows a twist of the nanowire from [100] to [120]. The TEM analysis beyond both thicker and ultra-thin TPS nanowires indicates an extremely high level of crystallinity, which has been successfully preserved during the LPE process.

More interestingly, we didn't observe significant oxidized or amorphous layer on the exfoliated nanowires even after storing in ambient environment for more than one month. This phenomenon is consistent with our calculations of $H_2O$ and $O_2$ adsorptions on TPS surface, as shown by Supplementary Figure 9. For the TPS crystal, the adsorption energy for $O_2$ molecule is 1.12 eV/molecule, indicating a completely not favorable adsorption of $O_2$. Although the TPS surface shows favorable adsorption of $H_2O$ molecule with an adsorption energy of -0.33 eV/molecule, the TPS structural geometry is stable upon such adsorption as indicated by Supplementary Figure 9. In fact, this stable crystal structure even at nanometer scale is crucial for modern electronic applications. Thus, we further characterize the device performance based on the high quality TPS nanowires.

**Functionality demonstrations and zero-bias photoresponse**

It has been demonstrated that the mechanical exfoliated TPS nanowire is a good candidate for the channel material of 1D field-effect transistor.[11] In order to study the electronic transport properties of TPS nanowires through LPE method, we fabricated FETs with a double-gate layout. As indicated by Figure 4a, a DC current is flowing through TPS nanowire which is tuned by a back gate and a top ionic liquid gate (Ionic liquid N,N-diethyl-N-(2-methoxyethyl)-N-methylammonium bis (trifluoromethylsulphonyl) imide (DEME-TFSI) has been used as the gate medium).[46] Figure 4b shows the transconductance for a typical FET based on a 31.4 nm TPS nanowire as indicated by the AFM image (Figure 4b inset). Under the back gate, at a source-drain bias of 0.5 V, the TPS nanowire clearly show n-type semiconducting behavior. The device can be switched to on-state with $I_{DS}$ = 5.1×10$^{-8}$ A at $V_{bg}$=80 V, and turned into off-state with $I_{DS}$ < 0.1 nA at $V_{bg}$ = -50 V, resulting in an on/off ratio larger than 500. Based on the back gate sweeps, the field effect mobility $\mu_{FE}$ can be extracted from the standard FET model $\mu_{FE} = [dI_{DS}/dV_{bg}] \times [L/(WC_{ox}V_{DS})]$, where $L$ is the channel length which is 6.7 μm, $W$ is the channel width as 31.4 nm. $C_{ox}$ is the capacitance per unit area which can be determined from $C_{ox} = \varepsilon_r \varepsilon_0 / d_{ox}$, where $\varepsilon_0$ is the dielectric constant of vacuum, $\varepsilon_r$ represents the relative dielectric constant of 3.9 for SiO$_2$, and $d_{ox}$ is 300nm as the dielectric layer thickness. For this specific device, $\mu_{FE}$ equals to 32.5 cm$^2$V$^{-1}$s$^{-1}$ at $V_{bg}$ = 80 V, which is comparable with that achieved through mechanical exfoliated TPS nanowire (with an average of 80 cm$^2$V$^{-1}$s$^{-1}$). Our TPS nanowires exhibit comparable or even better performance compared with other solution processed semiconducting nanostructures, i.e. MoS$_2$,[47] black phosphorus,[26] and carbon nanotubes.[33,48] To further improve the gate response of the TPS FETs, a top ionic liquid gate is applied in a range of -2 to 2 V. The device shows significant bipolar behavior with an on-current of $4.5 \times 10^{-7}$ A at the electron side and $1.4 \times 10^{-8}$ A at the hole side. The off-state current is as low as 10$^{-10}$ A, indicating a high on/off ratio near 10$^4$. The swing slope extracted from the gate sweeps has been improved from 23 V/dec for back gate to 517 mV/dec for ionic liquid gate. Figure 4c shows the linear $I_{DS}$-$V_{DS}$ dependence at different liquid gate voltages from -2 to 2 V, indicating nearly ohmic contacts between TPS nanowire and the gold metals.

Then, we further checked the stability of our TPS nanowires from the transport analysis. As shown in Supplementary Figure 10, the 1D TPS transistor shows very good stability with almost unchanged transconductance after stored in ambient environment for more than two months. We believe that the oxidation process is quite slow since it will increase the resistance or make the contacts worse significantly. This result is in good consistency with our STEM studies and theoretical calculations as discussed above. They have simultaneously provided solid supports for the good stability of our TPS nanostructures.

From the above analysis, our exfoliated TPS nanowires exhibit outstanding semiconducting properties. In previous studies with 1D or quasi-1D semiconductors, such as silicon nanowires,[49] carbon nanotubes,[50,51] ZnO nanowires,[52] and VO$_2$ nanobeams,[53,54] their promising applications as photodetectors have been addressed. Our following discussions will focus on the photo-response study based on TPS nano-devices. Figure 5a shows the optical image of a two-terminal photodetector device. Under our confocal microscope system, the 532 nm green laser (~1 mW) as the excitation light was focused at the center of the device. A typical time-resolved photoresponse is plotted in Figure 5b, $I_{photo}$ at zero source drain bias rises upon turning on the laser and decays after switching off the laser. As shown by the rise and decay process in Figure 5c, the response curve can be characterized by single exponential function expressed as $I_{photo} = A_0^* \exp\left(-\frac{t-t_0}{\tau}\right) + B_0$, where $t_0$ is the time when the laser is switched on, $A_0$ and $B_0$ are fitting constants, and $\tau$ is the time constant. The fitted time constants for rise ($\tau_R$) and decay ($\tau_D$) processes are 110ms and 117ms respectively. The response speed is slower than photovoltaic dominated behavior as reported in TMDCs[55-58] and their p-n junctions,[59,60] indicating a different mechanism in our TPS photodectector.

In order to explore the photocurrent generation mechanism, we presented scanning photocurrent microscopy measurements of our TPS device (see Methods). Figure 5d shows the 2D contour plot of the photocurrent depending on the laser spot locations at the device. The photocurrent exhibits a sign switching when the laser spot is scanned from source to drain electrodes, with maximum magnitudes at these metal-semiconductor contact edges. Besides, when the laser spot is located micrometer away from these two edges, either in-between or at the two sides, the photocurrent is still existing but with smaller magnitude. This is in controversy with Schottky barrier dominated photocurrent generation which is supposed to be constrained at the metal-semiconductor interfaces. Thus, we propose a photothermal dominated mechanism in our TPS 1D photodetector, as illustrated by Figure 5f. In particular, a local temperature gradient induced by light absorption drives electrons to travel through the device. When the laser is heating up the contacts, the Seebeck coefficient difference between Au metal and TPS nanowire generates a photothermal voltage across the junction, which could explain the maximum photocurrent at the contacts.[61-64] Meanwhile, as shown in Figure 5e and Supplementary Figure 11, the photocurrent profile along the line-cut from the red to blue dashed circles in Figure 5d does not show gate dependence, which is another clear evidence for the photothermal dominated mechanism.[64,65]

**CONCLUSION**

In summary, we have demonstrated the large-quantity production of high quality ultra-thin TPS nanowires using liquid-phase exfoliation with selected solvents. The resultant TPS nanowires show highly preserved crystallinity and can be considered as an important step toward the fabrication of perfect 1D TMDC nanoribbons with naturally semiconducting properties. 1D field effect transistors based on exfoliated TPS nanowires exhibit competitive electrical transport properties and high mobility among liquid exfoliated van der Waals semiconductors. Besides, the as fabricated devices show significant photoresponse with a clear photothermal dominated mechanism. Our work sheds light on the extension of wet exfoliation technique to 1D or quasi 1D van der Waals materials, which contains a large group of candidates. The efficient production of 1D semiconducting TPS nanowires can provide a wide range of applications, referred to as 1D field-effect transistor, optoelectronics, energy harvesting, chemical sensing, and biological sensing, etc.

**METHODS**

**Bulk Synthesis.**

Bulk Ta$_2$Pd$_3$Se$_8$ single crystal is synthesized using Chemical Vapor Transport method. The mixture of thoroughly ground stoichiometric Ta, Pd, and Se element powders was sealed in a quartz tube under vaccum and pre-heated for a week at 750 °C. The resultant powder was ground again and sealed in a new tube together with 75 mg iodine (transport agent). Then the tube was put into a double zone furnace with the charge end and the other end heated up to 850 °C and 900 °C respectively. The temperature was reversed with charge end at 900 °C and cold end at 850 °C after one day, and the reaction lasted for one week. After the furnace shut down, long needle-like single crystals with silver luster were formed at the cold end.

**Sample Preparations.**

Part A, liquid form sample preparations. 0.5 mg bulk TPS single crystal was immersed in 20 mL solvents (IPA or NMP). The mixture was sonicated for 4 hours at a bath temperature of 50 °C. The obtained dispersion was centrifuged at 500 rpm for 1 hour and the top half of the centrifuged solution was collected as the liquid form sample.

Part B, on-chip sample preparations. The on-chip samples were prepared using drop-casting methods. The substrate is Si/SiO$_2$ (300 nm) wafer which has been cut into 1 cm$^2$ square. The solution was drop cast on the substrate and dried by using a spin-coater to spin at 1000 rpm for 10 seconds. Then, the samples were rinsed in IPA for 2 mins and dried by using compressed air.

**Characterizations.**

Raman scattering spectroscopy measurements were carried out at room temperature using a micro-Raman spectrometer (Horiba-JY T64000) equipped with liquid nitrogen cooled charge-coupled device (CCD). The measurements were conducted in a backscattering configuration under a triple subtractive mode, excited with a diode green laser ($\lambda$ = 532 nm). We used a reflecting Bragg grating followed by another ruled reflecting grating to filter out the laser side bands, and as such we can achieve ∼5 cm$^{-1}$ limit of detection. The backscattered signal was collected through a 100× objective and dispersed by an 1800 g/mm grating with a spectra resolution of 0.7 cm$^{-1}$. For polarization measurement, we fix the analyzer before detector and the sample, and rotate the polarization direction of the laser excitation.

SEM was performed using Hitachi S-4800 FESEM. The images under all magnifications were acquired at 5 kV acceleration voltage. 1 nm thick carbon layer was uniformly deposited on the samples to reduce the charging effect caused by the SiO$_2$ substrate.

Atomic resolution ADF-STEM images of TPS nanowires were obtained using a JEOL annular field detector with a fine-imaging probe. The probe current was approximate 23 pA with a convergence semi-angle of 22 mrad and an inner semi-angle of 45-50 mrad.

**Mass fraction and yield estimations.**

The mass fraction of the exfoliated Ta$_2$Pd$_3$Se$_8$ nanowires can be estimated according to our AFM length and thickness distributions analysis. Given that the volume of individual nanowire is assumed as $V = \rho D^2 L$, where D is the diameter (thickness), L is the length, and $\rho$ is the density, the mass fraction of individual thickness can be estimated as:

$$F = \frac{\sum_{individual} D_{ind}^2 L_{ind}}{\sum_{all} D_{all}^2 L_{all}}$$

Then, since the sediment after centrifugation (500 rpm, 1 hour) has been weight as ~21 wt% of the initial bulk material, which indicates ~79 wt% remaining in the supernatant used for the AFM analysis. The total yield from the initial bulk TPS crystals can be calculated as $Y = F \times 79\%$.

**Raman active vibration modes simulations.**

All calculations of atomic and electronic structure of Ta$_2$Pd$_3$Se$_8$ nanowires were performed using a density functional theory[1, 2] within the PBE-PAW approximation[3] with the periodic boundary conditions using a Vienna Ab-initio Simulation Package.[4–6] The plane-wave energy cutoff was equal to 300 eV. To calculate the equilibrium atomic structures, the Brillouin zone was sampled according to the Monkhorst–Pack[7] scheme with a grid not less than 4×6×10 k-point for bulk structure and 1×1×12 k-point for quasi 1D nanowires structures. To avoid the interaction between the neighboring graphene layers, the vacuum space between them was greater than 15 Å. The structural relaxation was performed until the forces acting on each atom were less 0.01 eV/Å.

The Raman spectra spectrum simulation was performed using python script vasp_raman.py,[8] based on computation the derivative of the polarizability (or macroscopic dielectric tensor) with respect to that normal mode coordinate: dP/dQ (or de/dQ).[9]

**Photocurrent measurement setup.**

The photoresponse measurements were performed based on a confocal microscope setup. A ~1mW 532nm green laser beam was focused onto a ~1 μm$^2$ spot using a 50× microscope objective lens. The sample was mounted on an Attocube scanning stage with a scanning resolution of 100 nm both along x and y directions. The photocurrent at zero source drain bias was measured using a DC current amplifier and recorded by home-designed Labview software.

**SUPPLEMENTARY MATERIALS**

**REFERENCES**

(1). Novoselov, K. S., Geim, A. K., Morozov, S. V., Jiang, D., Zhang, Y., Dubonos, S. V., Grigorieva, I. V. & Firsov, A. A. Electric field effect in atomically thin carbon films. *Science* **306**, 666-669 (2004).
(2). Novoselov, K. S., Jiang, Z., Zhang, Y., Morozov, S. V., Stormer, H. L., Zeitler, U., Maan, J. C., Boebinger, G. S., Kim, P. & Geim, A. K. Room-temperature quantum hall effect in graphene. *Science* **315**, 1379-1379 (2007).
(3). Geim, A. K. & Novoselov, K. S. The rise of graphene. *Nat Mater* **6**, 183-191 (2007).
(4). Dean, C. R., Young, A. F., Meric, I., Lee, C., Wang, L., Sorgenfrei, S., Watanabe, K., Taniguchi, T., Kim, P., Shepard, K. L. & Hone, J. Boron nitride substrates for high-quality graphene electronics. *Nat Nanotechnol* **5**, 722-726 (2010).
(5). Novoselov, K. S., Jiang, D., Schedin, F., Booth, T. J., Khotkevich, V. V., Morozov, S. V. & Geim, A. K. Two-dimensional atomic crystals. *P Natl Acad Sci USA* **102**, 10451-10453 (2005).
(6). Radisavljevic, B., Radenovic, A., Brivio, J., Giacometti, V. & Kis, A. Single-layer MoS2 transistors. *Nat Nanotechnol* **6**, 147-150 (2011).
(7). Das, S., Chen, H. Y., Penumatcha, A. V. & Appenzeller, J. High Performance Multilayer MoS2 Transistors with Scandium Contacts. *Nano Lett* **13**, 100-105 (2013).
(8). Ovchinnikov, D., Allain, A., Huang, Y. S., Dumcenco, D. & Kis, A. Electrical Transport Properties of Single-Layer WS2. *Acs Nano* **8**, 8174-8181 (2014).
(9). Liu, X., Hu, J., Yue, C. L., Della Fera, N., Ling, Y., Mao, Z. Q. & Wei, J. High Performance Field-Effect Transistor Based on Multi layer Tungsten Disulfide. *Acs Nano* **8**, 10396-10402 (2014).
(10). Li, L. K., Yu, Y. J., Ye, G. J., Ge, Q. Q., Ou, X. D., Wu, H., Feng, D. L., Chen, X. H. & Zhang, Y. B. Black phosphorus field-effect transistors. *Nat Nanotechnol* **9**, 372-377 (2014).
(11). Liu, X., Liu, J. Y., Antipina, L. Y., Hu, J., Yue, C. L., Sanchez, A. M., Sorokin, P. B., Mao, Z. Q. & Wei, J. Direct Fabrication of Functional Ultrathin Single-Crystal Nanowires from Quasi-One-Dimensional van der Waals Crystals. *Nano Lett* **16**, 6188-6195 (2016).
(12). Stolyarov, M. A., Liu, G. X., Bloodgood, M. A., Aytan, E., Jiang, C. L., Samnakay, R., Salguero, T. T., Nika, D. L., Rumyantsev, S. L., Shur, M. S., Bozhilov, K. N. & Balandin, A. A. Breakdown current density in h-BN-capped quasi-1D TaSe3 metallic nanowires: prospects of interconnect applications. *Nanoscale* **8**, 15774-15782 (2016).
(13). Liu, G. X., Rumyantsev, S., Bloodgood, M. A., Salguero, T. T., Shur, M. & Balandin, A. A. Low-Frequency Electronic Noise in Quasi-1D TaSe3 van der Waals Nanowires. *Nano Lett* **17**, 377-383 (2017).
(14). Peng, B., Xu, K., Zhang, H., Ning, Z. Y., Shao, H. Z., Ni, G., Li, J., Zhu, Y. Y., Zhu, H. Y. & Soukoulis, C. M. 1D SbSeI, SbSI, and SbSBr With High Stability and Novel Properties for Microelectronic, Optoelectronic, and Thermoelectric Applications. *Advanced Theory and Simulations* **1**, 1700005 (2018).
(15). Geremew, A., Bloodgood, M. A., Aytan, E., Woo, B. W. K., Corber, S. R., Liu, G., Bozhilov, K. N., Salguero, T. T., Rumyantsev, S., Rao, M. P. & Balandin, A. A. Current carrying capacity of quasi-1D ZrTe3 van der Waals nanoribbons. *IEEE Electron Device Letters* **39**, 735-738 (2018).
(16). Bloodgood, M. A., Wei, P., Aytan, E., Bozhilov, K. N., Balandin, A. A. & Salguero, T. T. Monoclinic structures of niobium trisulfide. *APL Materials* **6**, 026602 (2018).
(17). Geremew, A., Kargar, F., Zhang, E. X., Zhao, S. E., Aytan, E., Bloodgood, M. A., Salguero, T. T., Rumyantsev, S., Fedoseyev, A., Fleetwood, D. M. & Balandin, A. A. Proton-irradiation-immune electronics implemented with two-dimensional charge-density-wave devices. *Nanoscale* **11**, 8380-8386 (2019).
(18). Fox, D., Zhou, Y., Maguire, P., Neill, A., Coileain, C., Gatensby, R., Glushenkov, A., Tao, T., Duesberg, G., Shvets, I. V., Abid, M., Wu, H., Chen, Y., Coleman, J. N., Donegan, J. F. & Zhang, H. Nanopatterning and Electrical Tuning of MoS2 Layers with a Subnanometer Helium Ion Beam. *Nano Lett* **15**, 5307-5313 (2015).
(19). Stanford, M., Pudasaini, P., Cross, N., Mahady, K., Hoffman, A., Mandrus, D., Duscher, G., Chisholm, M. & Rack, P. Tungsten Diselenide Patterning and Nanoribbon Formation by Gas-Assisted Focused Helium Ion Beam Induced Etching. *Small Methods*, 1600060 (2017).
(20). Nethravathi, C., Jeffery, A., Rajamathi, M., Kawamoto, N., Tenne, R., Golberg, D. & Bando, Y. Chemical Unzipping of WS2 nanotubes. *Acs Nano* **7**, 7311-7317 (2013).
(21). Lin, J., Peng, Z., Wang, G., Zakhidov, D., Rodriguez, E., Yacaman, M. & Tour, J. Enhanced Electrocatalysis for Hydrogen Evolution Reactions from WS2 Nanoribbons. *Advanced Energy Materials* **4** (2014).
(22). Hernandez, Y., Nicolosi, V., Lotya, M., Blighe, F. M., Sun, Z. Y., De, S., McGovern, I. T., Holland, B., Byrne, M., Gun'ko, Y. K., Boland, J. J., Niraj, P., Duesberg, G., Krishnamurthy, S., Goodhue, R., Hutchison, J., Scardaci, V., Ferrari, A. C. & Coleman, J. N. High-yield production of graphene by liquid-phase exfoliation of graphite. *Nat Nanotechnol* **3**, 563-568 (2008).
(23). Coleman, J. N., Lotya, M., O'Neill, A., Bergin, S. D., King, P. J., Khan, U., Young, K., Gaucher, A., De, S., Smith, R. J., Shvets, I. V., Arora, S. K., Stanton, G., Kim, H. Y., Lee, K., Kim, G. T., Duesberg, G. S., Hallam, T., Boland, J. J., Wang, J. J., Donegan, J. F., Grunlan, J. C., Moriarty, G., Shmeliov, A., Nicholls, R. J., Perkins, J. M., Grieveson, E. M., Theuwissen, K., McComb, D. W., Nellist, P. D. & Nicolosi, V. Two-Dimensional Nanosheets Produced by Liquid Exfoliation of Layered Materials. *Science* **331**, 568-571 (2011).
(24). Smith, R. J., King, P. J., Lotya, M., Wirtz, C., Khan, U., De, S., O'Neill, A., Duesberg, G. S., Grunlan, J. C., Moriarty, G., Chen, J., Wang, J. Z., Minett, A. I., Nicolosi, V. & Coleman, J. N. Large-Scale Exfoliation of Inorganic Layered Compounds in Aqueous Surfactant Solutions. *Adv Mater* **23**, 3944-+ (2011).
(25). Brent, J. R., Savjani, N., Lewis, E. A., Haigh, S. J., Lewis, D. J. & O'Brien, P. Production of few-layer phosphorene by liquid exfoliation of black phosphorus. *Chem Commun* **50**, 13338-13341 (2014).
(26). Yasaei, P., Kumar, B., Foroozan, T., Wang, C. H., Asadi, M., Tuschel, D., Indacochea, J. E., Klie, R. F. & Salehi-Khojin, A. High-Quality Black Phosphorus Atomic Layers by Liquid-Phase Exfoliation. *Adv Mater* **27**, 1887-+ (2015).

(27). Zhi, C. Y., Bando, Y., Tang, C. C., Kuwahara, H. & Golberg, D. Large-Scale Fabrication of Boron Nitride Nanosheets and Their Utilization in Polymeric Composites with Improved Thermal and Mechanical Properties. *Adv Mater* **21**, 2889-+ (2009).
(28). Warner, J. H., Rummeli, M. H., Bachmatiuk, A. & Buchner, B. Atomic Resolution Imaging and Topography of Boron Nitride Sheets Produced by Chemical Exfoliation. *Acs Nano* **4**, 1299-1304 (2010).
(29). Kelly, A. G., Hallam, T., Backes, C., Harvey, A., Esmaeily, A. S., Godwin, I., Coelho, J., Nicolosi, V., Lauth, J., Kulkarni, A., Kinge, S., Siebbeles, L. D. A., Duesberg, G. S. & Coleman, J. N. All-printed thin-film transistors from networks of liquid-exfoliated nanosheets. *Science* **356**, 69-72 (2017).
(30). Liu, J., Casavant, M. J., Cox, M., Walters, D. A., Boul, P., Lu, W., Rimberg, A. J., Smith, K. A., Colbert, D. T. & Smalley, R. E. Controlled deposition of individual single-walled carbon nanotubes on chemically functionalized templates. *Chem Phys Lett* **303**, 125-129 (1999).
(31). Bergin, S. D., Nicolosi, V., Streich, P. V., Giordani, S., Sun, Z. Y., Windle, A. H., Ryan, P., Niraj, N. P. P., Wang, Z. T. T., Carpenter, L., Blau, W. J., Boland, J. J., Hamilton, J. P. & Coleman, J. N. Towards solutions of single-walled carbon nanotubes in common solvents. *Adv Mater* **20**, 1876-+ (2008).
(32). Coleman, J. N. Liquid-Phase Exfoliation of Nanotubes and Graphene. *Adv Funct Mater* **19**, 3680-3695 (2009).
(33). Cao, Q., Han, S. J., Tulevski, G. S., Franklin, A. D. & Haensch, W. Evaluation of field-effect mobility and contact resistance of transistors that use solution-processed single-walled carbon nanotubes. *Acs Nano* **6**, 6471-6477 (2012).
(34). Fuhrer, M. S., Kim, B. M., Durkop, T. & Brintlinger, T. High-mobility nanotube transistor memory. *Nano Lett* **2**, 755-759 (2002).
(35). Durkop, T., Getty, S. A., Cobas, E. & Fuhrer, M. S. Extraordinary mobility in semiconducting carbon nanotubes. *Nano Lett* **4**, 35-39 (2004).
(36). Kang, S. J., Kocabas, C., Ozel, T., Shim, M., Pimparkar, N., Alam, M. A., Rotkin, S. V. & Rogers, J. A. High-performance electronics using dense, perfectly aligned arrays of single-walled barbon nanotubes. *Nat Nanotechnol* **2**, 230-236 (2007).
(37). Cao, Q., Kim, H., Pimparkar, N., Kulkarni, J. P., Wang, C. J., Shim, M., Roy, K., Alam, M. A. & Rogers, J. A. Medium-scale carbon nanotube thin-film integrated circuits on flexible plastic substrates. *Nature* **454**, 495-500 (2008).
(38). Cui, Y., Zhang, Z., Wang, D., Wang, W. U. & Lieber, C. M. High performance silicon nanowire field effect transistors. *Nano Lett* **3**, 149-152 (2003).
(39). Duan, X., Niu, C., Sahi, V., Chen, J., Parce, J. W., Empedocles, S. & Goldman, J. L. High-performance thin-film transistors using semiconductor nanowires and nanoribbons. *Nature* **425**, 274-278 (2003).
(40). Tang, M. S., Ng, E. P., Juan, J. C., Ooi, C. W., Ling, T. C., Woon, K. L. & Show, P. L. Metallic and semiconducting carbon nanotubes separation using an aqueous two-phase separation technique: a review. *Nanotechnology* **27**, 332002 (2016).
(41). Zhao, Y., Luo, X., Li, H., Zhang, J., Araujo, P. T., Gan, C. K., Wu, J., Zhang, H., Quek, S. Y., Dresselhaus, M. S. & Xiong, Q. H. Interlayer breathing and shear modes in few-trilayer MoS2 and WSe2. *Nano Lett* **13**, 1007-1015 (2013).
(42). Claus, R., Hacker, H. H., Schrotter, H. W., Brandmuller, J. & Haussuhl, S. Low-Frequency Optical-Phonon Spectrum Of Benzil. *Phys Rev* **187**, 1128-+ (1969).
(43). Ren, Z. Q., McNeil, L. E., Liu, S. B. & Kloc, C. Molecular motion and mobility in an organic single crystal: Raman study and model. *Phys Rev B* **80** (2009).
(44). Ye, H. Q., Liu, G. F., Liu, S., Casanova, D., Ye, X., Tao, X. T., Zhang, Q. C. & Xiong, Q. H. Molecular-Barrier-Enhanced Aromatic Fluorophores in Cocrystals with Unity Quantum Efficiency. *Angew Chem Int Edit* **57**, 1928-1932 (2018).

(45). Kim, J., Lee, J. U., Lee, J., Park, H. J., Lee, Z., Lee, C. & Cheong, H. Anomalous polarization dependence of Raman scattering and crystallographic orientation of black phosphorus. *Nanoscale* **7**, 18708-18715 (2015).
(46). Braga, D., Lezama, I. G., Berger, H. & Morpurgo, A. F. Quantitative Determination of the Band Gap of WS2 with Ambipolar Ionic Liquid-Gated Transistors. *Nano Lett* **12**, 5218-5223 (2012).
(47). Lee, K., Kim, H. Y., Lotya, M., Coleman, J. N., Kim, G. T. & Duesberg, G. S. Electrical Characteristics of Molybdenum Disulfide Flakes Produced by Liquid Exfoliation. *Adv Mater* **23**, 4178-+ (2011).
(48). Kim, W. J., Lee, C. Y., O'Brien, K. P., Plombon, J. J., Blackwell, J. M. & Strano, M. S. Connecting single molecule electrical measurements to ensemble spectroscopic properties for quantification of single-walled carbon nanotube separation. *J. Am. Chem. Soc.* **131**, 3128-3129 (2009).
(49). Ahn, Y., Dunning, J. & Park, J. Scanning photocurrent imaging and electronic band studies in silicon nanowire field effect transistors. *Nano Lett* **5**, 1367-1370 (2005).
(50). Freitag, M., Tsang, J. C., Bol, A., Avouris, P., Yuan, D. N. & Liu, J. Scanning photovoltage microscopy of potential modulations in carbon nanotubes. *Appl Phys Lett* **91** (2007).
(51). Avouris, P., Freitag, M. & Perebeinos, V. Carbon-nanotube photonics and optoelectronics. *Nat Photonics* **2**, 341-350 (2008).
(52). Kind, H., Yan, H. Q., Messer, B., Law, M. & Yang, P. D. Nanowire ultraviolet photodetectors and optical switches. *Adv Mater* **14**, 158-+ (2002).
(53). Li, Z. J., Hu, Z. P., Peng, J., Wu, C. Z., Yang, Y. C., Feng, F., Gao, P., Yang, J. L. & Xie, Y. Ultrahigh Infrared Photoresponse from Core-Shell Single-Domain-VO2/V2O5 Heterostructure in Nanobeam. *Adv Funct Mater* **24**, 1821-1830 (2014).
(54). Wu, J. M. & Chang, W. E. Ultrahigh Responsivity and External Quantum Efficiency of an Ultraviolet-Light Photodetector Based on a Single VO2 Microwire. *Acs Appl Mater Inter* **6**, 14286-14292 (2014).
(55). Lopez-Sanchez, O., Lembke, D., Kayci, M., Radenovic, A. & Kis, A. Ultrasensitive photodetectors based on monolayer MoS2. *Nat Nanotechnol* **8**, 497-501 (2013).
(56). Yin, Z. Y., Li, H., Li, H., Jiang, L., Shi, Y. M., Sun, Y. H., Lu, G., Zhang, Q., Chen, X. D. & Zhang, H. Single-Layer MoS2 Phototransistors. *Acs Nano* **6**, 74-80 (2012).
(57). Zhang, W., Chiu, M. H., Chen, C. H., Chen, W., Li, L. J. & Wee, A. T. S. Role of Metal Contacts in High-Performance Phototransistors Based on WSe2 Monolayers. *Acs Nano* **8**, 8653-8661 (2014).
(58). Octon, T. J., Nagareddy, V. K., Russo, S., Craciun, M. F. & Wright, C. D. Fast High-Responsivity Few-Layer MoTe2 Photodetectors. *Adv Opt Mater* **4**, 1750-1754 (2016).
(59). Furchi, M. M., Pospischil, A., Libisch, F., Burgdorfer, J. & Mueller, T. Photovoltaic Effect in an Electrically Tunable van der Waals Heterojunction. *Nano Lett* **14**, 4785-4791 (2014).
(60). Lee, C. H., Lee, G. H., van der Zande, A. M., Chen, W. C., Li, Y. L., Han, M. Y., Cui, X., Arefe, G., Nuckolls, C., Heinz, T. F., Guo, J., Hone, J. & Kim, P. Atomically thin p-n junctions with van der Waals heterointerfaces. *Nat Nanotechnol* **9**, 676-681 (2014).
(61). Balasubramanian, K., Fan, Y. W., Burghard, M., Kern, K., Friedrich, M., Wannek, U. & Mews, A. Photoelectronic transport imaging of individual semiconducting carbon nanotubes. *Appl Phys Lett* **84**, 2400-2402 (2004).
(62). Tsen, A. W., Donev, L. A. K., Kurt, H., Herman, L. H. & Park, J. Imaging the electrical conductance of individual carbon nanotubes with photothermal current microscopy. *Nat Nanotechnol* **4**, 108-113 (2009).
(63). Buscema, M., Barkelid, M., Zwiller, V., van der Zant, H. S. J., Steele, G. A. & Castellanos-Gomez, A. Large and Tunable Photothermoelectric Effect in Single-Layer MoS2. *Nano Lett* **13**, 358-363 (2013).
(64). Buchs, G., Bagiante, S. & Steele, G. A. Identifying signatures of photothermal current in a double-gated semiconducting nanotube (vol 5, 4987, 2014). *Nat Commun* **6** (2015).
(65). Ahn, Y. H., Tsen, A. W., Kim, B., Park, Y. W. & Park, J. Photocurrent imaging of p-n junctions in ambipolar carbon nanotube transistors. *Nano Lett* **7**, 3320-3323 (2007).

**Methods Reference**

(1). Hohenberg, P. & Kohn, W. Inhomogeneous electron gas. *Phys. Rev.* **136**, 864-871 (1964).
(2). Kohn, W. & Sham, L. J. Self-consistent equations including exchange and correlation effects. *Phys. Rev.* **140**, 1133-1138 (1965).
(3). Perdew, J. P., Burke, K. & Ernzerhof, M. Generalized Gradient Approximation Made Simple. *Phys. Rev. Lett.* **77**, 3865-3868 (1996).
(4). Kresse, G. & Hafner, J. Ab initio molecular dynamics for liquid metals. *Phys. Rev. B* **47**, 558–561 (1993).
(5). Kresse, G. & Hafner, J. Ab initio molecular-dynamics simulation of the liquid-metal-amorphous-semiconductor transition in germanium. *Phys. Rev. B* **49**, 14251–14269 (1994).
(6). Kresse, G. & Furthmüller, J. Efficient iterative schemes for ab initio total-energy calculations using a plane-wave basis set. *Phys. Rev. B* **54**, 11169-11186 (1996).
(7). Monkhorst, H. J. & Pack, J. D. Special points for Brillouin-zone integrations. *Phys. Rev. B* **13**, 5188-5192 (1976).
(8). Fonari, A. & Stauffer, S. *vasp_raman.py*. (https://github.com/raman-sc/VASP/, 2013).
(9). Porezag, D. & Pederson, M. R. Infrared intensities and Raman-scattering activities within density-functional theory. *Phys. Rev. B* **54**, 7830-7836 (1996).

**Acknowledgements**

This work is supported by the United States Department of Energy under Grant DE-SC0014208 and by The National Science Foundation under Grant 1752997. We acknowledge the Coordinated Instrument Facility (CIF) of Tulane University for the support of various instruments. P.B.S. and L.Y.A. (theoretical calculations) were supported by Russian Science Foundation (Project identifier: 17-72-20223). We are grateful to the supercomputer cluster provided by the Materials Modelling and Development Laboratory at NUST "MISIS" (supported via the Grant from the Ministry of Education and Science of the Russian Federation No. 14.Y26.31.0005), and to the Joint Supercomputer Center of the Russian Academy of Sciences.

**Author contributions**

X.L. and J.W. conceived and designed this project. J.L. and Z.M. synthesized and characterized the bulk crystals. X.L. performed the exfoliations. X.L., S.L., Y.Z., C.Y., A.J., A.M.S., and Q.X. performed microscopy characterizations. X.L. fabricated the devices, performed electrical transport and photoresponse measurements. P.B.S., L.Y.A., J.N., and J.S. performed theoretical calculations. The manuscript was written by X.L., and revised by X.L. and J.W. with input from all other authors. This project was supervised by J.W.

**Competing financial interests**

The authors declare no competing financial interest.

**FIGURES**

**Figure 1. Crystal structure and exfoliation results.** **(a)** Stereo-view of $Ta_2Pd_3Se_8$ bulk crystal structure along *c*-axis. The unit ribbon is highlighted by the dashed rectangular. **(b)** Simulated $Ta_2Pd_3Se_8$ nano-structures achieved by ultra-sonication exfoliations with the assistance of chemical solvents (IPA). **(c)** $Ta_2Pd_3Se_8$ dispersions in IPA and NMP solvents right after sonication. **(d)** Resultant liquids after 1 hour's centrifugation at 500 rpm for IPA and NMP. **(e)**, **(f)** Scanning Electron Microscopy images of $Ta_2Pd_3Se_8$ nanowires deposited on Si wafer using liquids from (c) and (d), respectively. Inset, SEM zoom-in view of the selected 15 μm×15 μm square.

**Figure 2. Basic characterizations of exfoliated nanowires.** **(a), (b)** Atomic Force Microscopy images of the on-chip deposited $Ta_2Pd_3Se_8$ nanowires exfoliated using (a) IPA and (b) NMP solvents, respectively. Insets in the zoom-in view of selected areas (dashed blue rectangular 5 μm × 2.5 μm) show height profiles of four individual nanowires, indicating thicknesses of 2.2 nm, 1.4 nm, 4.6 nm, and 1.7 nm. **(c), (d)** Contour plots of the thickness and length distributions of the $Ta_2Pd_3Se_8$ nanowires produced by (c) IPA and (d) NMP solvents, respectively. **(e)** Raman spectra of a thick (~1 μm) TPS fiber under different excitation polarizations. 0º and 90º denote parallel and cross polarization configurations respectively. Note, the simulated Raman spectrum has been plotted at the bottom for comparison, the low frequency part from 0 to 100 $cm^{-1}$ has been multiplied by a factor of 10. **(f)** Raman intensity polarization dependence for P1 (32.5 $cm^{-1}$), P2 (72.9 $cm^{-1}$), and P3 (201.5 $cm^{-1}$), respectively. Scattered points: measured peak intensity. Solid lines: guide lines for the polarization patterns.

**Figure 3. High resolution Scanning Transmission Electron Microscopy studies.** **(a)** Annular dark-field (ADF) transmission electron microscopy images of a thicker (9.3 nm) $Ta_2Pd_3Se_8$ nanowire along the <100>

zone axis. From left to right, ADF images of the same nanowire with increasing magnifications show the uniformity and smooth surface. Top right inset: fast Fourier transform pattern of the selected area. **(b)** ADF images of an ultra-thin (3.2 nm) $Ta_2Pd_3Se_8$ nanowire. Left: low magnification image shows the nanowire curvature. Right: atomic resolution images of three selected areas. Note: simulated crystal structures have been fitted perfectly with STEM images.

**Figure 4. Transport properties for the fabricated 1D-transistors. (a)** Schematic drawing for the fabricated $Ta_2Pd_3Se_8$ nanowire field-effect transistor with a double-gate setup. **(b)** Transconductance gate dependence for a typical field-effect transistor based on a 31.4 nm nanowire. Red and black dots are linked to top (ionic liquid gate voltage) and bottom (back gate voltage) axis, respectively. Inset: atomic force microscopy image of the measured nanowire device. **(c)** $I_{DS}$-$V_{DS}$ sweeps at different liquid gate voltages, 2V, 1.6V, 1.2V, 0V, -1.2V, and -2V.

**Figure 5. Photoresponse study based on $Ta_2Pd_3Se_8$ transistors. (a)** Optical image of the fabricated nanowire device and schematic drawing for the photocurrent mapping setup. The laser used in this experiment is 532nm solid laser. **(b)** Room-temperature photoresponse with a few light on/off cycles based on the device in (a). Note: the source-drain bias has been set to zero. **(c)** Zoom-in view of one cycle showing the rise and decay processes. Black dots: measured photo-response curve. Red solid line: fitted curve with exponential decay function. **(d)** Photocurrent mapping contour plot based on the device in (a). **(e)** Gate dependent photocurrent profiles across the line-cut starting from red dashed circle to the blue dashed circle in (d), from left to right, the back gate voltage is set to -60V, 0V, and 60V. **(f)** Band diagram of the photothermalelectric mechanism.

Figure 1

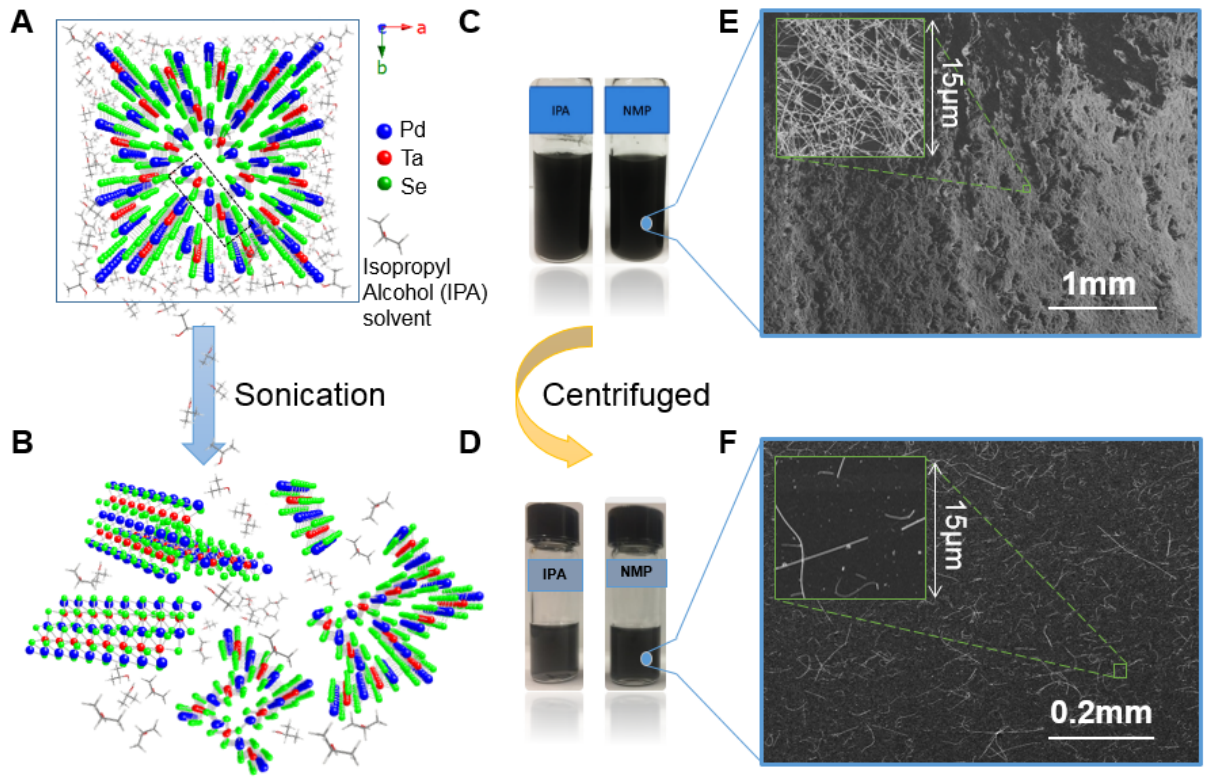

Figure 2

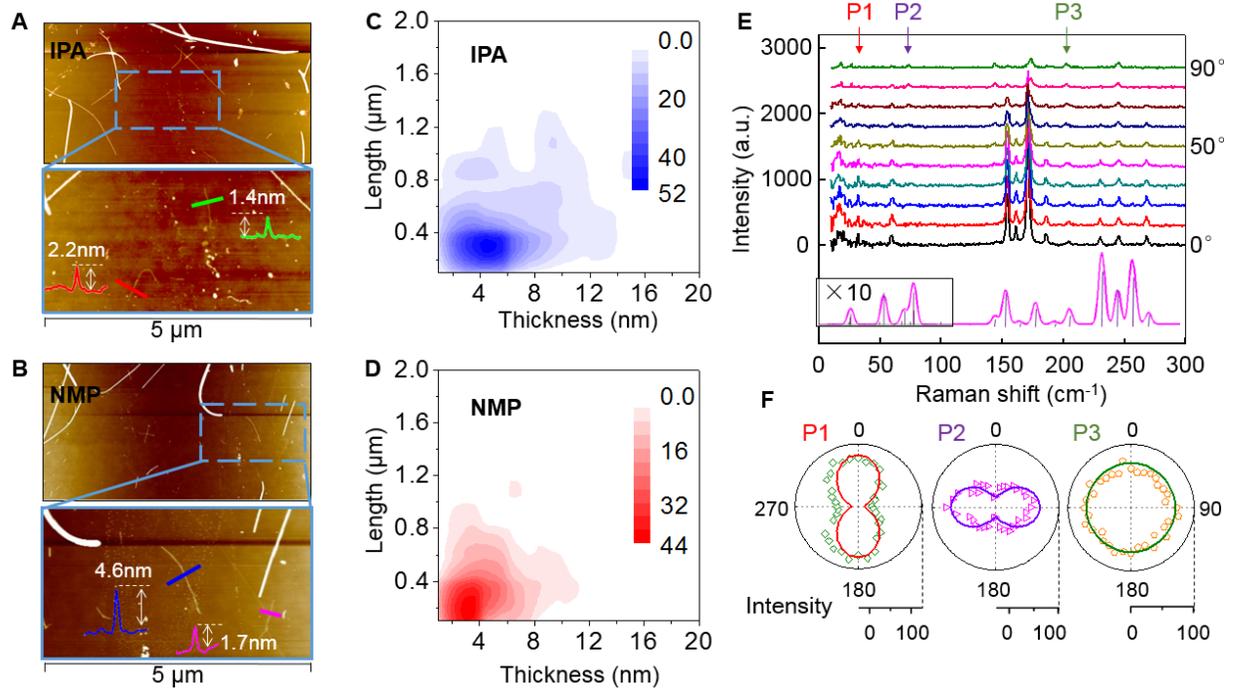

Figure 3

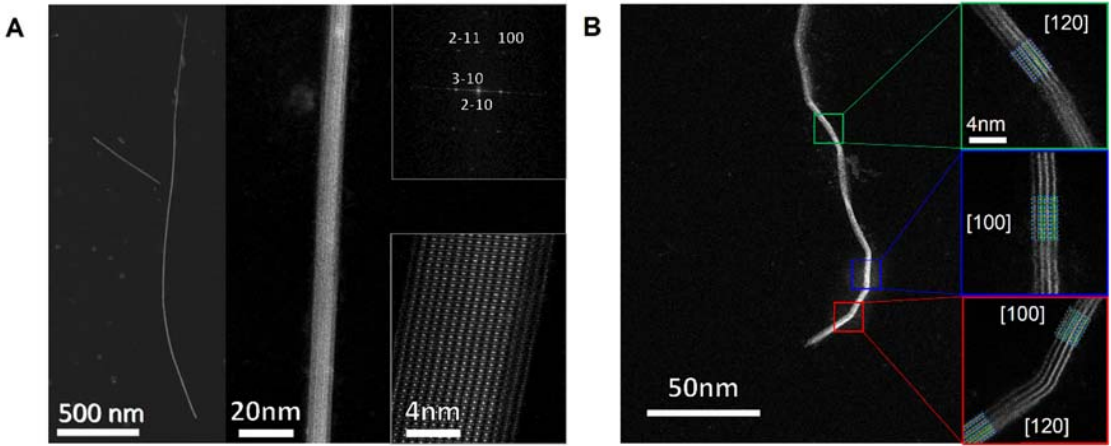

Figure 4

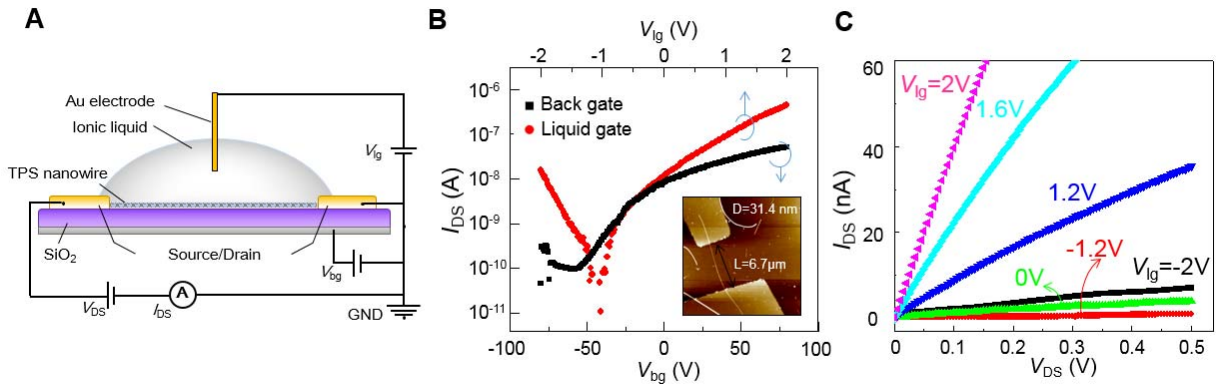

Figure 5

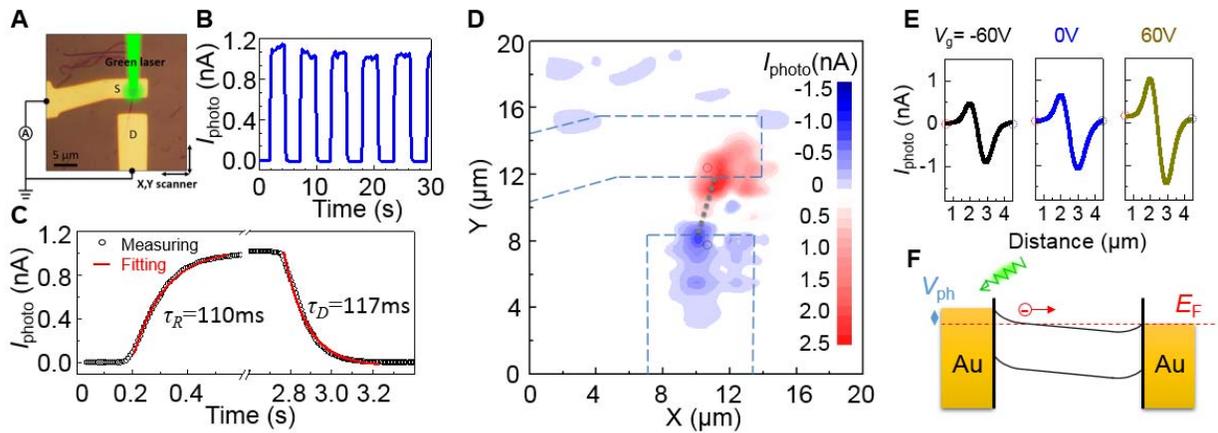